\documentclass[12pt,aps,prd,tightenlines,showkeys,superscriptaddress]{revtex4}

\usepackage{epsfig,floatflt}
\usepackage{amssymb,amsmath,citesort,indentfirst,fancyhdr,enumerate}
\usepackage{rotating}

\graphicspath{{ps}}

\def\nima#1#2#3{{Nucl.\ Instr.\ and Meth.} {\bf A#1}, #3 (#2)}

\def\npb#1#2#3{{ Nucl.\ Phys.}             {\bf B#1}, #3 (#2)}

\def\prd#1#2#3{{ Phys.\ Rev.}              {\bf D#1}, #3 (#2)}
\def\pr#1#2#3 {{ Phys.\ Rev.}              {\bf  #1}, #3 (#2)}

\def\be{\begin{equation}}
\def\ee{\end{equation}}
\def\bea{\begin{eqnarray}}
\def\eea{\end{eqnarray}}

\def\bckpp{B^\pm\to K^\pm\pi^\pm\pi^\mp}

\def\bpkpp{B^+\to K^+\pi^+\pi^-}
\def\bpkkk{B^+\to K^+K^+K^-}
\def\bnkpp{B^0\to K^0\pi^+\pi^-}
\def\kckk{K^\pm K^+K^-}
\def\kcpp{K^\pm\pi^\pm\pi^\mp}
\def\kppp{K^+\pi^+\pi^-}

\def\kspp{K^0_S\pi^+\pi^-}
\def\kpkk{K^+K^+K^-}

\def\de{\Delta E}
\def\mb{M_{bc}}

\def\qqbar{q\bar{q}}
\def\chic{\chi_{c0}}
\def\kpkm{K^+K^-}

\def\pipi{\pi^+\pi^-}
\def\pipi{\pi^+\pi^-}
\def\ks{K^0_S}
\def\BF{{\cal{B}}}

\def\ACP{A_{CP}}

\begin{document}

\title{Dalitz Analysis of $B\to Khh$ Decays at Belle}


\affiliation{Princeton University, Princeton, New Jersey 08544, U.S.A.}
\author{Alexey Garmash}\affiliation{Princeton University, Princeton, New Jersey 08544, U.S.A.}
\collaboration{Representing The Belle Collaboration}

\begin{abstract}
We report results on the Dalitz analysis of three-body charmless $\bpkpp$,
$\bnkpp$ and $\bpkkk$ decays including searches for direct $CP$ violation in
the $\bpkpp$ mode. Branching fractions for a number of quasi-two-body
intermediate states are reported. We also observe evidence with $3.9\sigma$ 
significance for a
large direct $CP$ violation in $B^\pm\to\rho(770)^0K^\pm$ channel.
This is the first evidence for $CP$ violation in a charged meson decay. The
results are obtained using a Dalitz analysis technique with a large data sample
of $B\bar{B}$ pairs collected with the Belle detector operating at the KEKB
asymmetric energy $e^+ e^-$ collider.
\end{abstract}

\maketitle

%%%%%%%%%%%%%%%%%%%%%%%%%%%%%%%%%%%%%%%%%%%%
%% MAINMATTER
%%%%%%%%%%%%%%%%%%%%%%%%%%%%%%%%%%%%%%%%%%%%

\section{Introduction}

Decays of $B$ mesons to three-body charmless hadronic final states provide
a rich laboratory for studying $B$ meson decay dynamics and provide new
possibilities for $CP$ violation searches. In decays to two-body
final states ($B\to K\pi$, $\pi\pi$, etc.) direct $CP$ violation can only be
observed as a difference in $B$ and $\bar{B}$ decay rates. In decays to
three-body final states dominated by quasi-two-body channels,
direct $CP$ violation can also manifest itself as a difference in relative
phases between two quasi-two-body channels. Large direct $CP$ violation is
expected in charged $B$ decays to some quasi-two-body charmless hadronic
modes~\cite{beneke-neubert}.

\section{Apparatus, Data Sample \& Event Selection}

The Dalitz analysis of $\bpkpp$ and $\bpkkk$ decays is performed with a
140\,fb$^{-1}$ data sample; for a Dalitz analysis of the $\bnkpp$ decay and
for $CP$ violation searches in the decay $\bpkpp$, we use a data sample of
357\,fb$^{-1}$. The data are collected with the Belle detector~\cite{Belle}
operating at the KEKB asymmetric-energy $e^+e^-$ collider with a
center-of-mass (c.m.) energy at the $\Upsilon(4S)$ resonance.

\begin{figure}[!t]
 \includegraphics[width=0.33\textwidth]{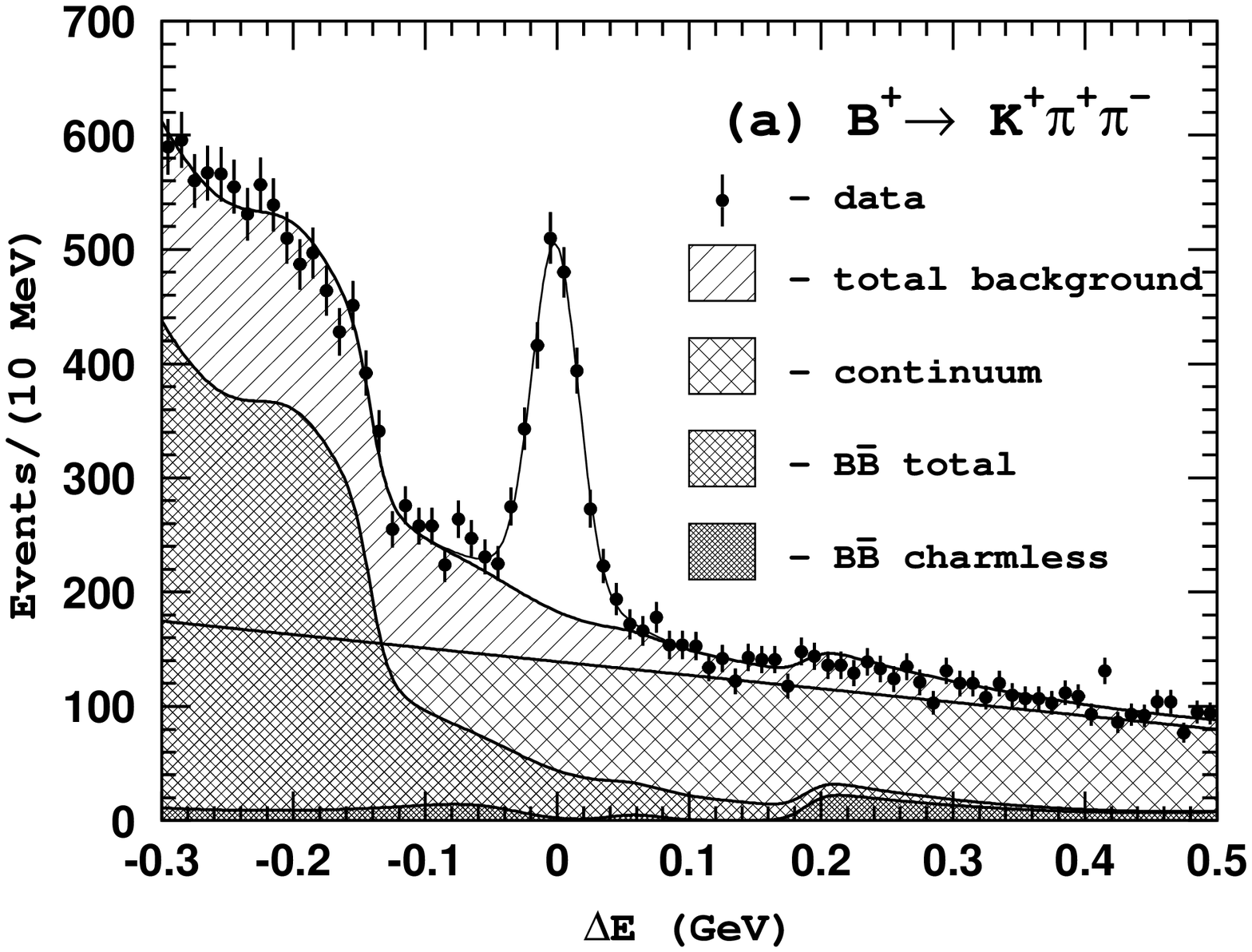} \hspace*{-2mm}
 \includegraphics[width=0.33\textwidth]{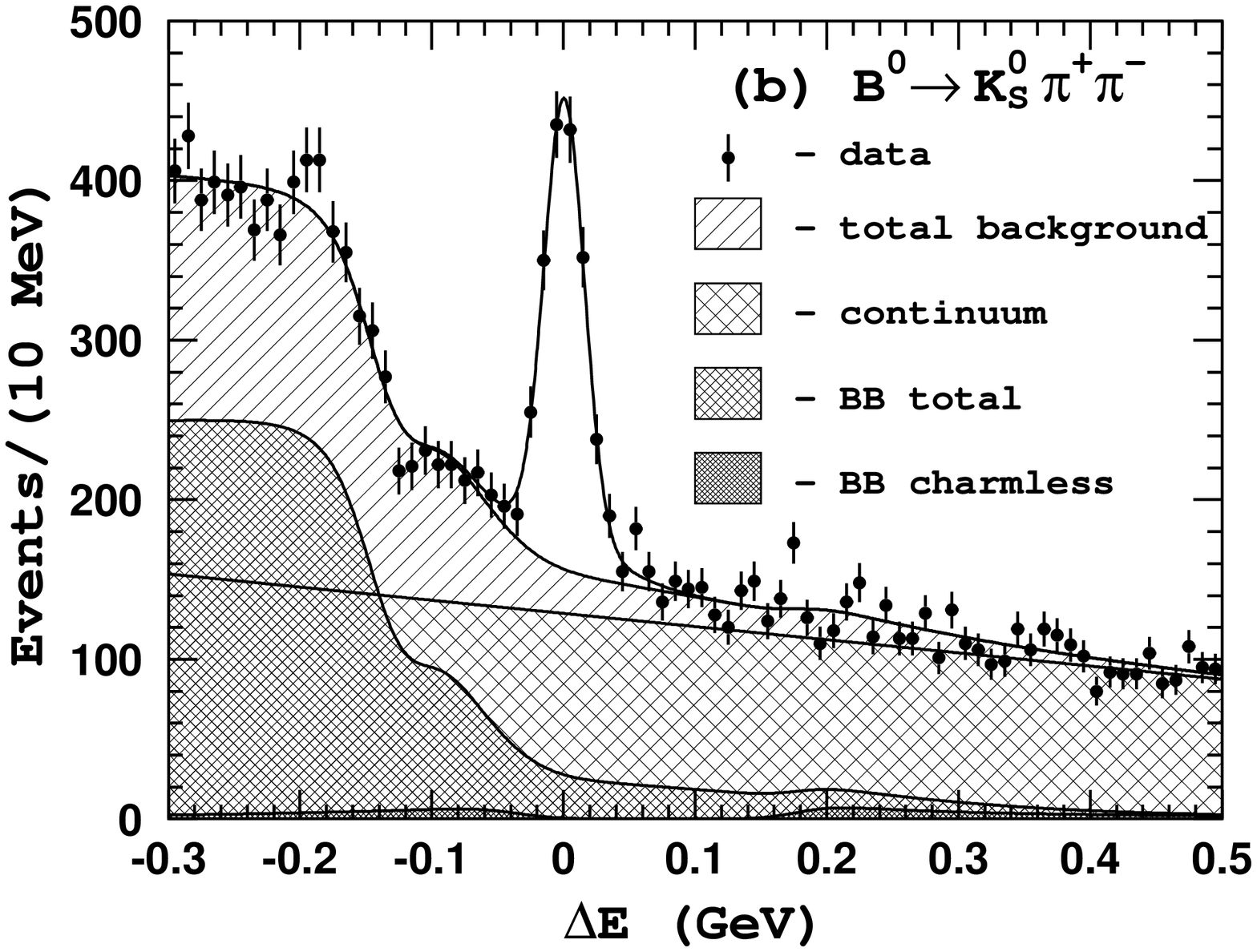} \hspace*{-2mm}
 \includegraphics[width=0.33\textwidth]{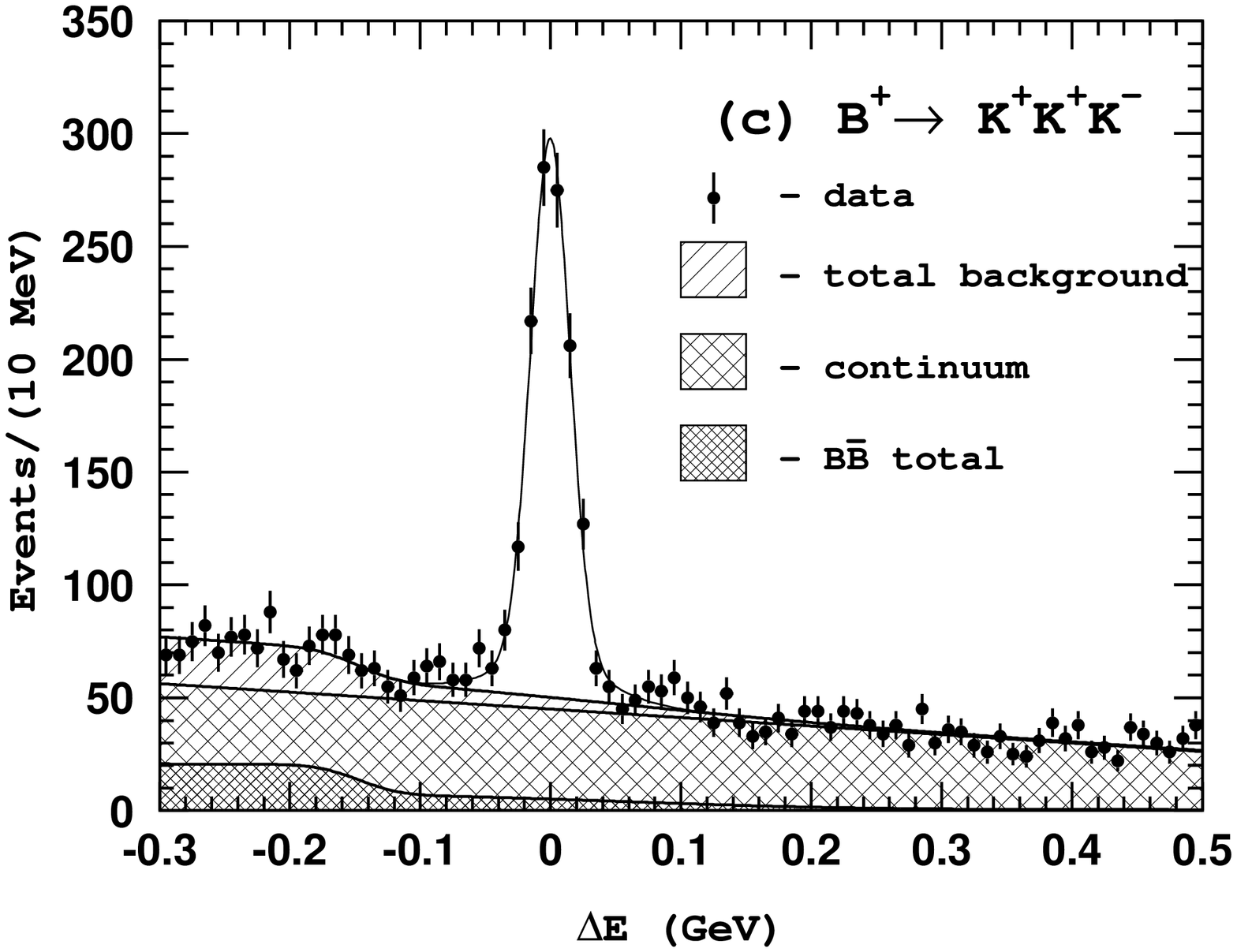}
 \caption{$\de$ distribution for the (a) $\bpkpp$, (b) $\bnkpp$ and
         (c) $\bpkkk$ candidate events with
          $|\mb-M_B|<7.5$~MeV/$c^2$. Points with error bars are data; the
          upper curve is the fit result; the hatched histograms are various
          background components.}
 \label{fig:dE}
\end{figure}

We identify $B$ candidates using two almost independent kinematic variables:
$\de = (\sum_i\sqrt{c^2{\bf p}_i^2 + c^4m_i^2} ) - E^*_{\rm beam}$ and 
$\mb =  \frac{1}{c^2}\sqrt{E^{*2}_{\rm beam}-c^2(\sum_i {\bf p}_i)^2},$
where the summation is over all particles from a $B$ candidate; ${\bf p}_i$
and $m_i$ are their c.m.\ three-momenta and masses, respectively.
The dominant background to studied processes is due to $e^+e^-\to~\qqbar$
($q=u,d,s$ and $c$ quarks) continuum events. This background is
 suppressed using variables that characterize the event topology. 
A detailed description of the continuum suppression technique can be found
in Ref.~\cite{khh-belle} and references therein. From a MC study we find
the dominant background originating from other $B$ decays that peaks in the
signal region is due to $B\to Dh$, where $h$ stands for a charged pion or
kaon and due to $B\to J/\psi(\psi(2S))[\mu^+\mu^-]K$ decays, where muons are
misidentified as pions. We veto these backgrounds by applying requirements on
the invariant mass of the appropriate two-particle combination. The most
significant background from charmless $B$ decays to $B\to K\pi\pi$ channels
originates from $B\to\eta'[\gamma\pipi]K$, from $B^+\to\pi^+\pi^+\pi^-$,
where one of the two same charge pions is misidentified as a kaon, and from
$B\to K\pi$ processes. These backgrounds cannot be removed and are taken into
account when fitting the data. We find no charmless $B$ decay modes that
produce a significant background to the $\kpkk$ final state. The $\de$
distributions for $\kppp$, $\kspp$ and $\kpkk$ candidates that pass all the
selection requirements are shown in Fig.~\ref{fig:dE}.

\section{Dalitz Analysis Results}

\begin{figure}[!t]
 \begin{tabular}{lcr}
  \includegraphics[width=0.33\textwidth]{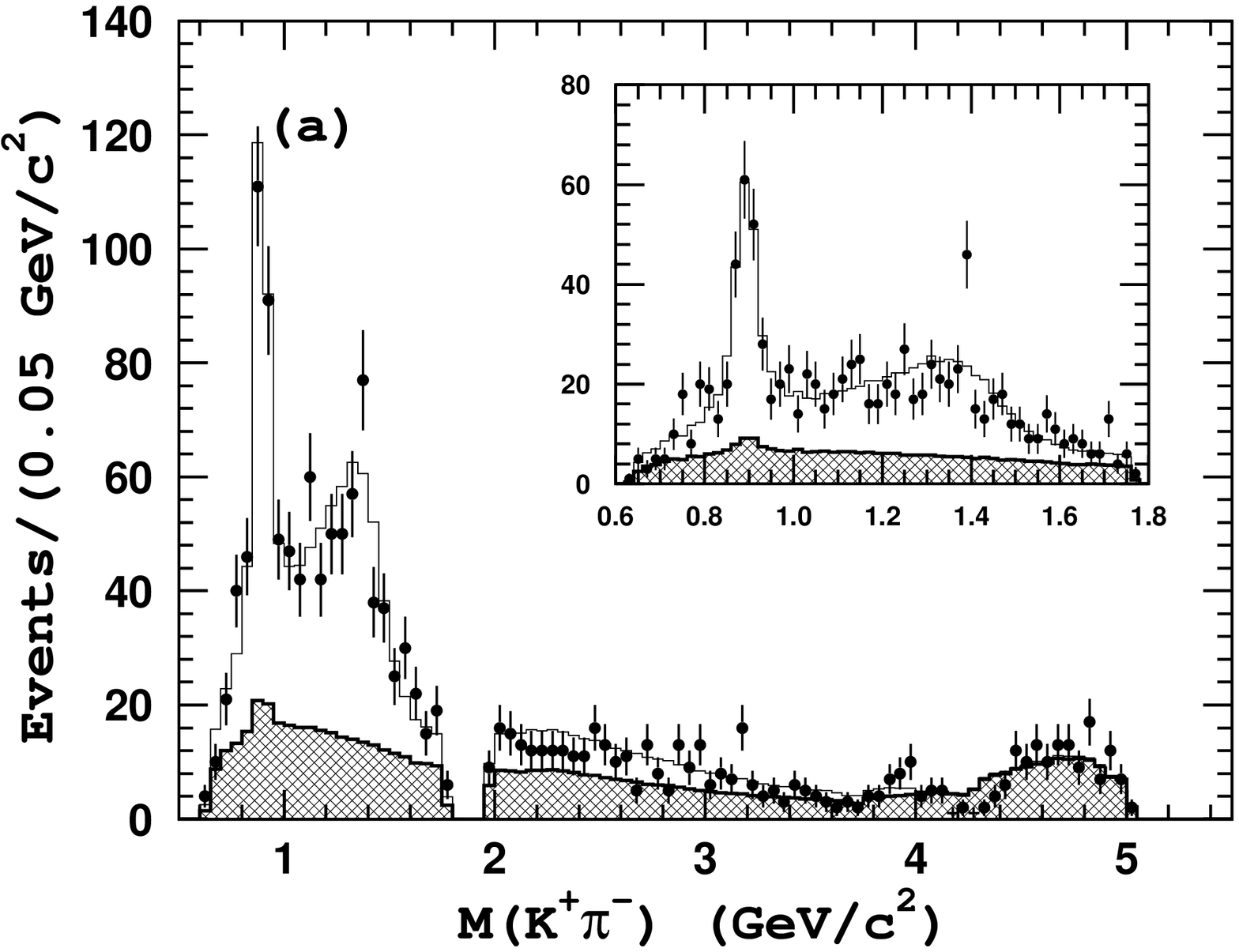}
  \includegraphics[width=0.33\textwidth]{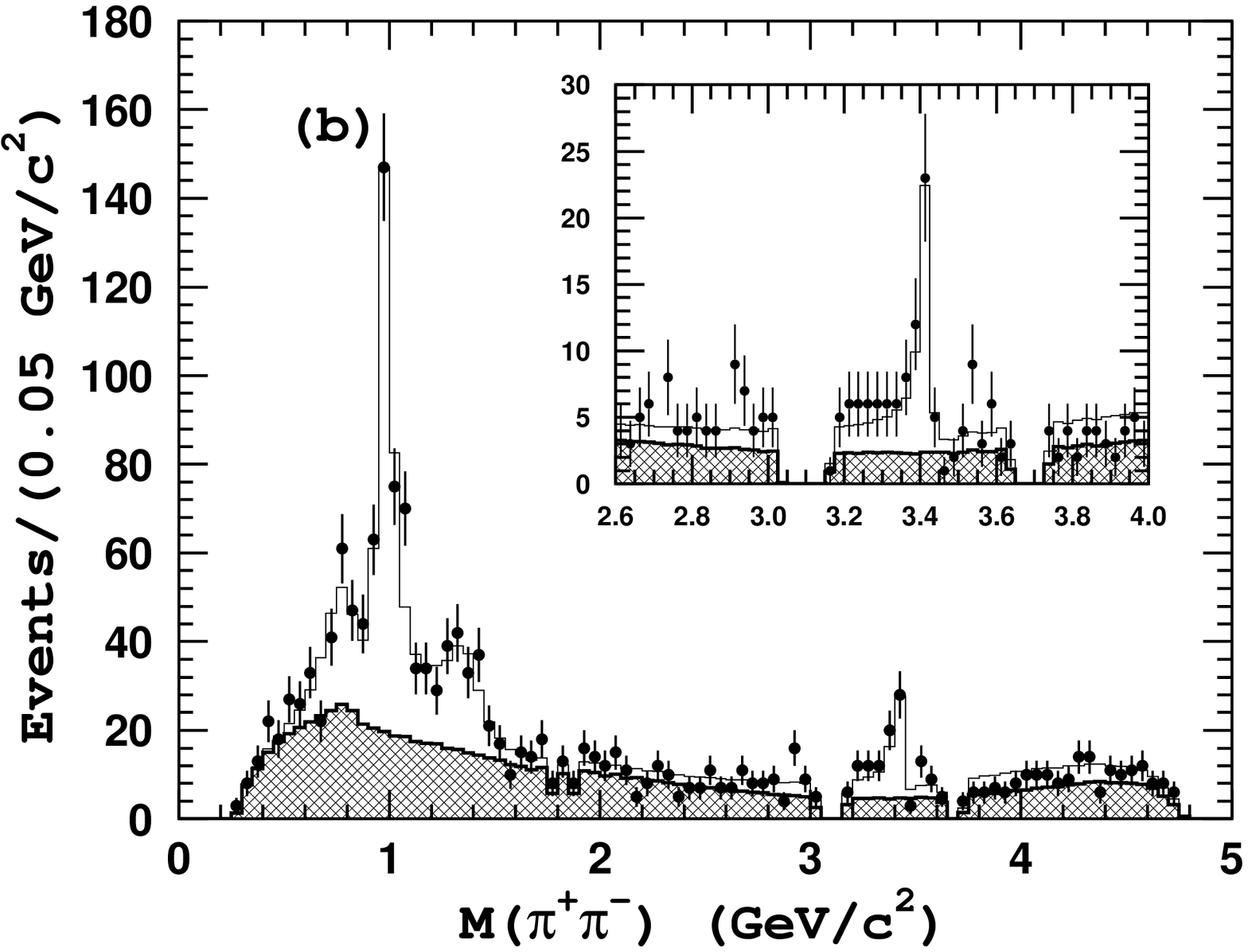}
  \includegraphics[width=0.33\textwidth]{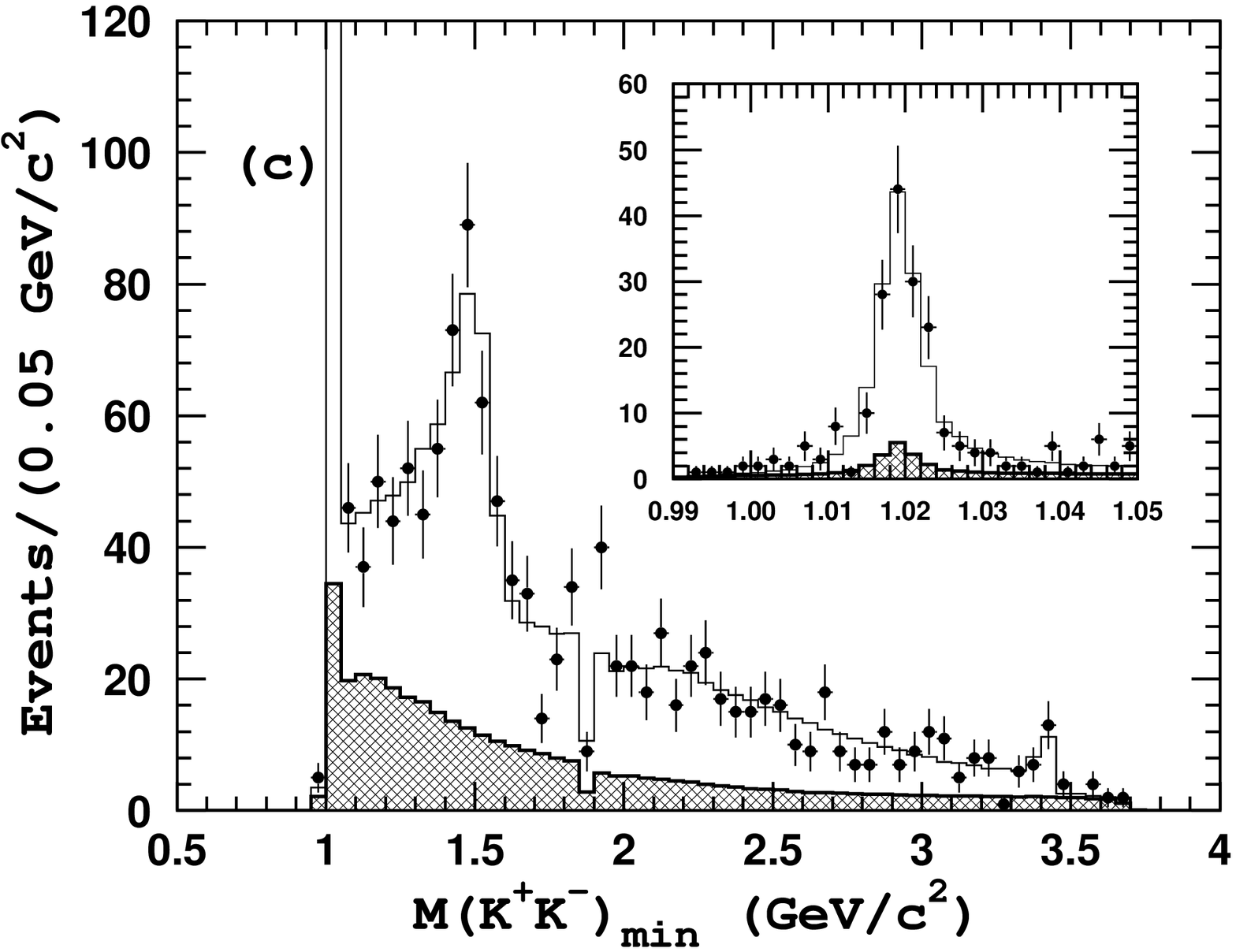}
\vspace*{-3mm} \\
  \includegraphics[width=0.33\textwidth]{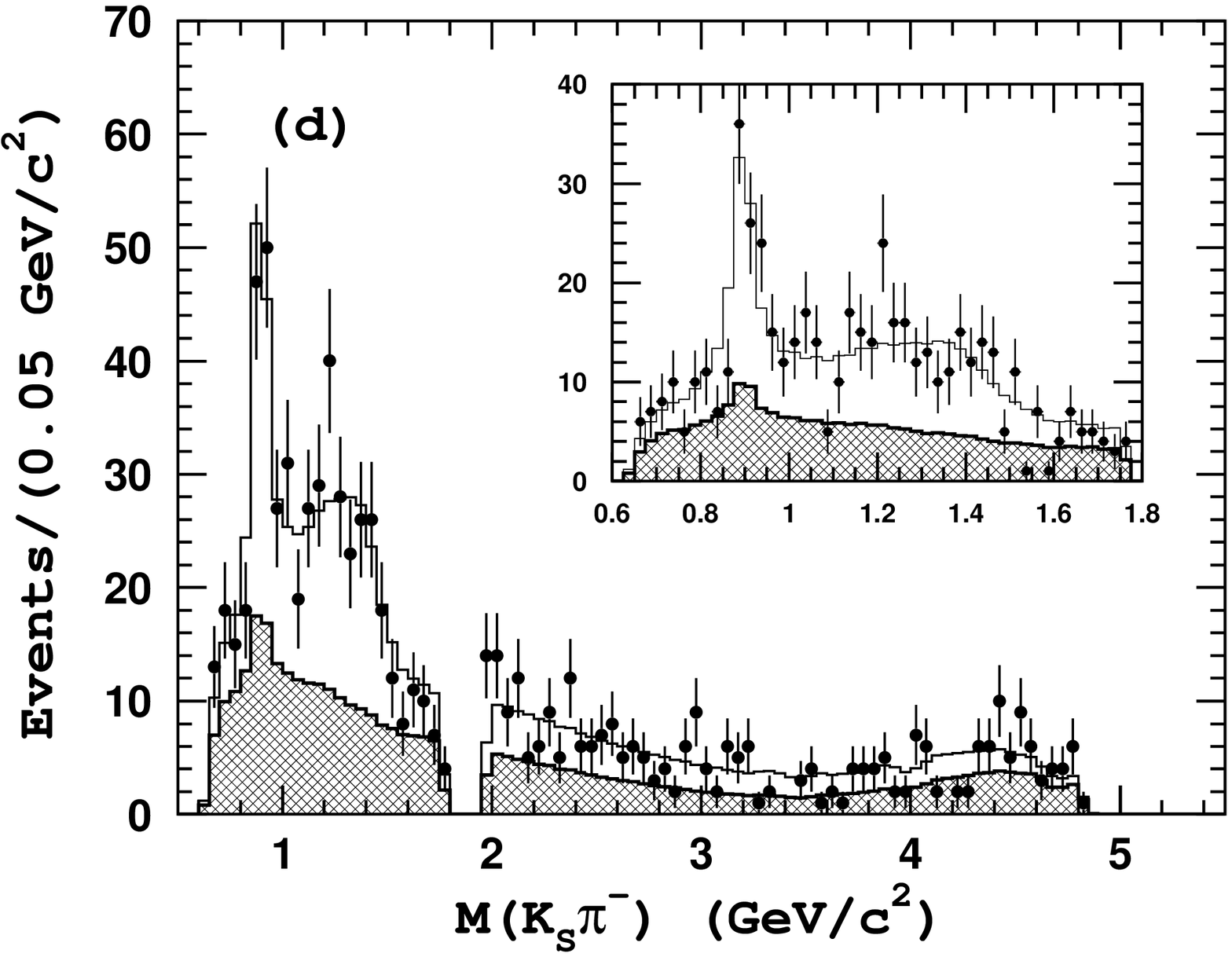}
  \includegraphics[width=0.33\textwidth]{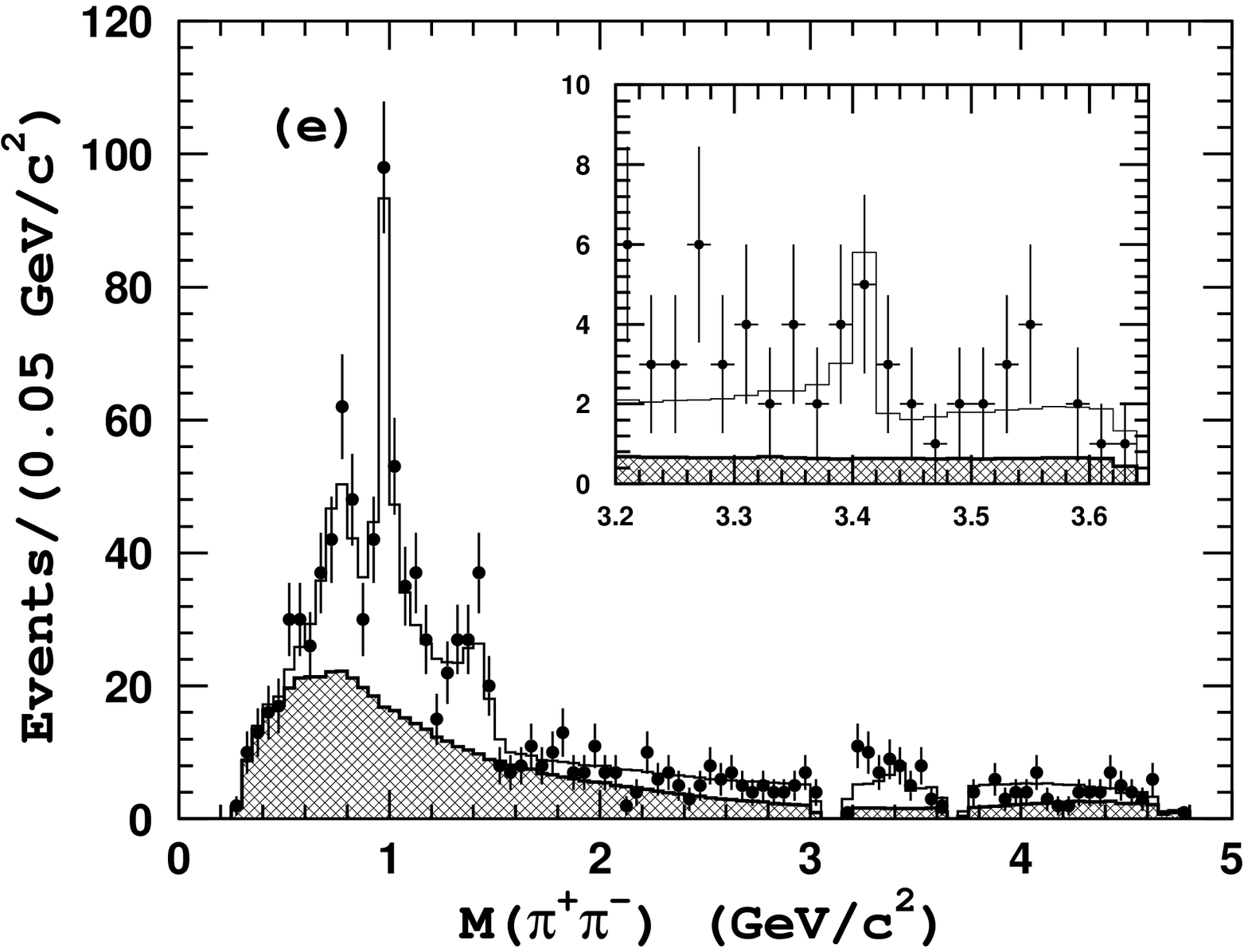}
  \includegraphics[width=0.33\textwidth]{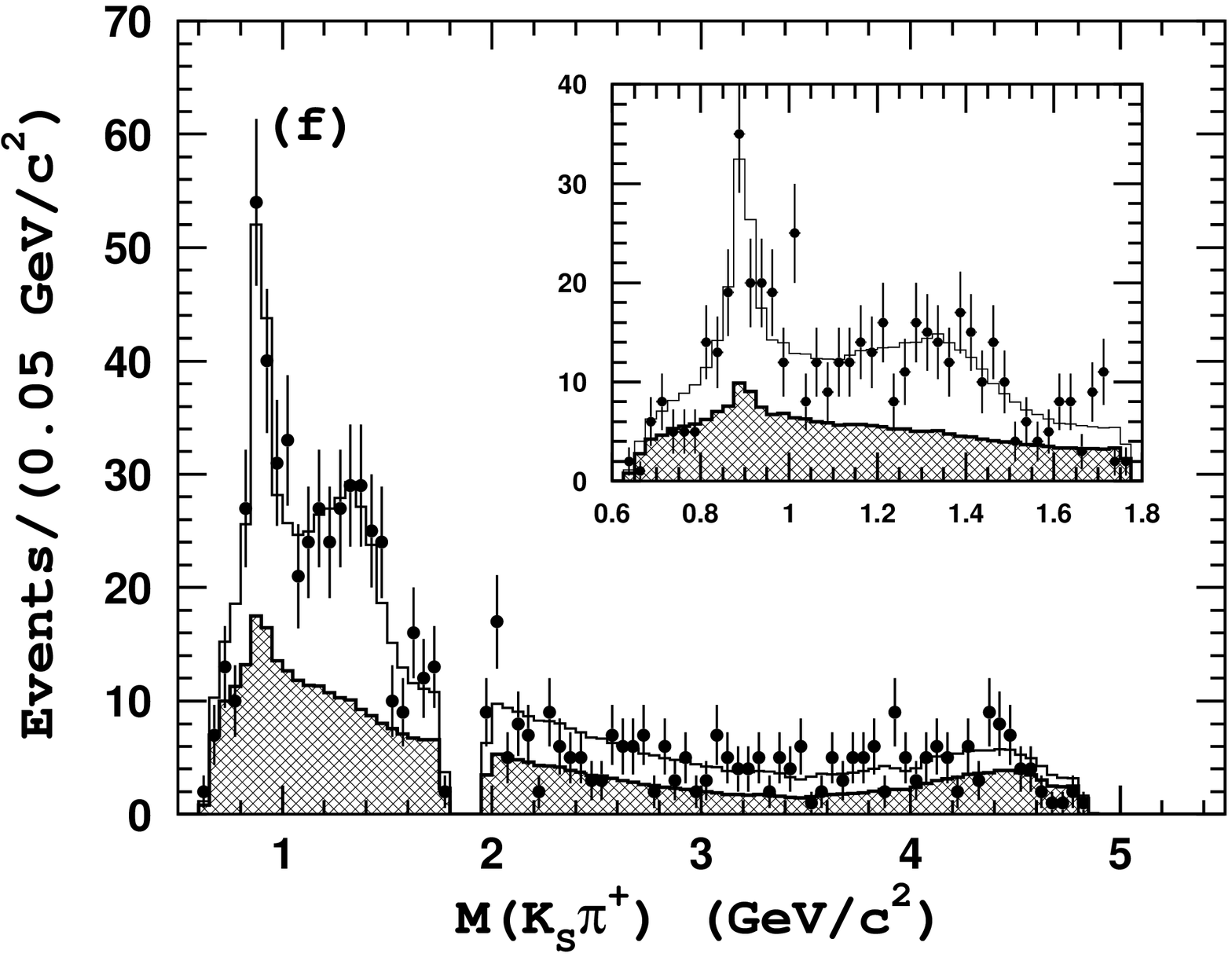}
\end{tabular}
\caption{Results of the fit to events in the $B$ signal region for the
         (a,b) $\kppp$, (c) $\kpkk$, and (d,e,f) $\kspp$ final state.
         Points with error bars are data, the open histogram is the fit
         result and hatched histogram is the background component.}
\label{fig:hh-mass}
\medskip
\end{figure}

For the amplitude analysis we select events in the $B$ signal region defined
as an ellipse around the $\mb$ and $\de$ mean values:
$\left[\frac{\mb-M_B}{7.5~{\rm MeV}/c^2}\right]^2+
 \left[\frac{\de}{40~{\rm MeV}}\right]^2<1.$ We find that $\bpkpp$ signal is
well described by a coherent sum of the $K^*(892)^0\pi^+$,
$K^*_0(1430)^0\pi^+$, $\rho(770)^0K^+$, $f_0(980)K^+$, $f_X(1300)K^+$,
$\chic K^+$ quasi-two-body channels and a non-resonant component. The channel
$f_X(1300)K^+$ (with mass and width of $f_X(1300)$ to be determined from the
fit) is added to account for an excess of signal events visible in $M(\pipi)$
spectrum near $1.3$~GeV/$c^2$. Results of the best fit are shown in
Figs.~\ref{fig:hh-mass} (a,b). The mass and width of the $f_X(1300)$ state
obtained from the fit are consistent with those for the $f_0(1370)$, however
more data are required for more definite conclusion. To test for the
contribution of other possible quasi-two-body intermediate states such as
$K^*(1410)^0\pi^+$, $K^*(1680)^0\pi^+$, $K^*_2(1430)^0\pi^+$ or $f_2(1270)K^+$,
we include an additional amplitude for each of these channels in the decay
amplitude one by one and repeat the fit to data. None of these channels have a
statistically significant signal. Branching fraction and upper limit results
are summarized in Table~\ref{tab:khh-branch}.
For more details see Ref.~\cite{khh-dalitz-belle}. 

\begin{table}[!t]
  \caption{Summary of branching fraction results. The first quoted error is
           statistical, the second is systematic and the third is the model
           error.}
  \medskip
  \label{tab:khh-branch}
\centering
  \begin{tabular}{lcr} \hline
\multicolumn{1}{c}{Mode} &
\hspace*{0mm}$\BF(B^+\to Rh^+)\times\BF(R\to h^+h^-)\times10^{6}$ &
\hspace*{7mm}$\BF(B^+\to Rh^+)\times10^{6}$  \\
 \hline
 $\kppp$ charmless total & $-$
                        & $46.6\pm2.1\pm4.3$  \\
 $~~K^*(892)^0[K^+\pi^-]\pi^+$
                        & $6.55\pm0.60\pm0.60^{+0.38}_{-0.57}$
                        & $9.83\pm0.90\pm0.90^{+0.57}_{-0.86}$     \\
 $~~K^*_0(1430)^0[K^+\pi^-]\pi^+$
                        & $27.9\pm1.8\pm2.6^{+8.5}_{-5.4}$
                        & $45.0\pm2.9\pm6.2^{+13.7}_{-~8.7}$       \\
 $~~K^*(1410)^0[K^+\pi^-]\pi^+$
                        & $<2.0$ & \multicolumn{1}{c}{$-$}        \\
 $~~K^*(1680)^0[K^+\pi^-]\pi^+$
                        & $<3.1$ & \multicolumn{1}{c}{$-$}        \\
 $~~K^*_2(1430)^0[K^+\pi^-]\pi^+$
                        & $<2.3$ & \multicolumn{1}{c}{$-$}        \\
 $~~\rho(770)^0[\pi^+\pi^-]K^+$
                        & $4.78\pm0.75\pm0.44^{+0.91}_{-0.87}$
                        & $4.78\pm0.75\pm0.44^{+0.91}_{-0.87}$     \\
 $~~f_0(980)[\pi^+\pi^-]K^+$
                        & $7.55\pm1.24\pm0.69^{+1.48}_{-0.96}$
                        & \multicolumn{1}{c}{$-$}                 \\
 $~~f_2(1270)[\pi^+\pi^-]K^+$
                        & $<1.3$ &\multicolumn{1}{c}{$-$}         \\
 ~~Non-resonant
                        & $-$
                        & $17.3\pm1.7\pm1.6^{+17.1}_{-7.8}$        \\
 $\kspp$  charmless     &  $-$
                        & $47.5\pm2.4\pm3.7$  \\
 $~~K^*(892)^+[K^0\pi^+]\pi^-$
                        & $5.61\pm0.72\pm0.43^{+0.43}_{-0.29}$
                        & $8.42\pm1.08\pm0.65^{+0.64}_{-0.43}$     \\
 $~~K^*_0(1430)^+[K^0\pi^+]\pi^-$
                        & $30.8\pm2.4\pm2.4^{+0.8}_{-3.0}$
                        & $49.7\pm3.8\pm3.8^{+1.2}_{-4.8}$         \\
 $~~K^*(1410)^+[K^0\pi^+]\pi^-$
                        & $<3.8$ & \multicolumn{1}{c}{$-$}         \\
 $~~K^*(1680)^+[K^0\pi^+]\pi^-$
                        & $<2.6$ & \multicolumn{1}{c}{$-$}         \\
 $~~K^*_2(1430)^+[K^0\pi^+]\pi^-$
                        & $<2.1$ & \multicolumn{1}{c}{$-$}         \\
 $~~\rho(770)^0[\pi^+\pi^-]K^0$
                        & $6.13\pm0.95\pm0.47^{+1.00}_{-1.05}$
                        & $6.13\pm0.95\pm0.47^{+1.00}_{-1.05}$     \\
 $~~f_0(980)[\pi^+\pi^-]K^0$
                        & $7.60\pm1.66\pm0.59^{+0.48}_{-0.67}$
                        & \multicolumn{1}{c}{$-$}                  \\
 $~~f_2(1270)[\pi^+\pi^-]K^0$
                        & $<1.4$ & \multicolumn{1}{c}{$-$}         \\
  ~~Non-resonant
                        & $-$
                        & $19.9\pm2.5\pm1.5^{+0.7}_{-1.2}$         \\
 $\kckk$ charmless total & $-$
                        & $30.6\pm1.2\pm2.3$                       \\
 $~~\phi [K^+K^-]K^+$
                        & $4.72\pm0.45\pm0.35^{+0.39}_{-0.22}$
                        & $9.60\pm0.92\pm0.71^{+0.78}_{-0.46}$     \\
 $~~\phi(1680)[K^+K^-]K^+$
                        & $<0.8$
                        & \multicolumn{1}{c}{$-$}                 \\
 $~~f_0(980)[K^+K^-]K^+$
                        & $<2.9$
                        & \multicolumn{1}{c}{$-$}                 \\
 $~~f'_2(1525)[K^+K^-]K^+$
                        & $<4.9$
                        & \multicolumn{1}{c}{$-$}                 \\
 $~~a_2(1320)[K^+K^-]K^+$
                        & $<1.1$
                        & \multicolumn{1}{c}{$-$}                 \\
 ~~Non-resonant
                        & $-$
                        & $24.0\pm1.5\pm1.8^{+1.9}_{-5.7}$         \\
\hline
 $\chic [\pi^+\pi^-]K^+$
                        & $1.37\pm0.28\pm0.12^{+0.34}_{-0.35}$
                        & \multicolumn{1}{c}{$-$}                 \\
 $\chic [K^+K^-]K^+$
                        & $0.86\pm0.26\pm0.06^{+0.20}_{-0.05}$
                        & \multicolumn{1}{c}{$-$}                 \\
 $\chic K^+$ combined   & $-$
                        & $196\pm35\pm33^{+197}_{-26}$             \\
 $\chic [\pi^+\pi^-]K^0$
                        & $<0.56$
                        & $<113$                                   \\
\hline
  \end{tabular}
\end{table}

In the fit to $\kspp$ events we use decay amplitude ${\cal{M}}$ similar to
those constructed in the analysis of $\bpkpp$ decay. The masses and widths of
all resonances are fixed at either their world average values from PDG or at
values determined from the analysis of $\bpkpp$ decay (for $f_0(980)$ and
$f_X(1300)$). Since in this analysis we do not distinguish between $B$ or
$\bar{B}$ decays the signal PDF is a non-coherent sum
$   S(\ks\pi^\pm\pi^\mp) = |{\cal{M}}(\ks\pi^+\pi^-)|^2 +
                           |{\cal{M}}(\ks\pi^-\pi^+)|^2.$
Results of the best fit are shown in Figs.~\ref{fig:hh-mass} (d-f). All plots
demonstrate good agreement between data and the fit. Branching fraction results
are summarized in Table~\ref{tab:khh-branch}.
For more details see Ref.~\cite{kspp-dalitz-belle}. 

We find that the $\kckk$ signal is well described by an amplitude that is a
coherent sum of the  $\phi K^+$, $f_X(1500)K^+$, $\chic K^+$ quasi-two-body
channels and a non-resonant component, where the $f_X(1500)K^+$ (with mass
and width of $f_X(1500)$ to be determined from the fit) channel is added to
describe the excess of signal events visible in $\kpkm$ mass spectrum near
$1.5$~GeV/$c^2$. As there are two identical kaons in the final state, the
decay amplitude is symmetrized with respect to interchange of two kaons of
the same charge
$S(K^+K^+K^-) = |{\cal{M}}(K^+_1K^+_2K^-) +
                 {\cal{M}}(K^+_2K^+_1K^-)|^2.$
Results of the fit are shown in Fig.~\ref{fig:hh-mass}~(c)
and summarized in Table~\ref{tab:khh-branch}. 
For more details see Ref.~\cite{khh-dalitz-belle}.

\section{Search for Direct $CP$ Violation in $\bckpp$}

%%%%%%%%%%%%%%%%%%%%%%%%%%%%%%%%%%%%%%%%%%%%%%%%%%%%%%%%%%%%%%%%%%%%%%%%%%%%%%%
\begin{table}[!t]
\caption{Results of the best fit to $\kcpp$ events in the $B$ signal region.
The first quoted error is statistical and the second is the model dependent
uncertainty. The quoted significance is statistical only.}
\medskip
\label{tab:kpp-dcpv}
\centering
  \begin{tabular}{lcccc}
\hline
          Channel
        & $b$ 
        & $\varphi$, ($^\circ$)
        & $\ACP$, (\%)
        & Significance, ($\sigma$)\\
\hline
$K^*(892)\pi^\pm$    
                     & $0.078\pm0.033^{+0.012}_{-0.003}$
                     & $ -18\pm44^{+5}_{-13}$
                     & $-14.9\pm6.4^{+0.8}_{-0.8}$ 
                     & $2.6$ \\
$K_0(1430)\pi^\pm$   
                     & $0.069\pm0.031^{+0.010}_{-0.008}$
                     & $-123\pm16^{+4}_{-5}$
                     & $+7.6\pm3.8^{+2.0}_{-0.9}$
                     & $2.7$ \\
$\rho(770)^0K^\pm$   
                     & $0.28\pm0.11^{+0.07}_{-0.09}$
                     & $-125\pm32^{+10}_{-85}$
                     & $+30\pm11^{+11}_{-4}$
                     & $3.9$ \\
$f_0(980)K^\pm$      
                     & $0.30\pm0.19^{+0.05}_{-0.10}$
                     & $-82\pm8^{+2}_{-2}$
                     & $-7.7\pm6.5^{+4.1}_{-1.6}$
                     & $1.6$ \\
$f_2(1270)K^\pm$     
                     & $0.37\pm0.17^{+0.11}_{-0.04}$
                     & $-24\pm29^{+14}_{-20}$
                     & $-59\pm22^{+3}_{-3}$
                     & $2.7$ \\
\hline
$\chic K^\pm$        
                     & $0.15\pm0.35^{+0.08}_{-0.07}$
                     & $-77\pm94^{+154}_{-11}$
                     & $-6.5\pm19.6^{+2.9}_{-1.4}$ 
                     & $0.7$ \\
\hline
  \end{tabular}
\end{table}
%%%%%%%%%%%%%%%%%%%%%%%%%%%%%%%%%%%%%%%%%%%%%%%%%%%%%%%%%%%%%%%%%%%%%%%%%%%%%%%

For $CP$ violation studies the amplitude for each quasi-two-body channel is
modified from $ae^{i\delta}$ to $ae^{i\delta}(1\pm be^{i\varphi})$, where the
plus (minus) sign corresponds to $B^+$ ($B^-$) decay. With such a
parameterization, the charge asymmetry, $\ACP$, for a particular quasi-two-body
$B\to f$ channel can be calculated as
\begin{equation}
A_{CP}(f)
      = \frac{N^--N^+}{N^-+N^+}
      = -\frac{2b\cos\varphi}{1+b^2}.
\label{eq:acp-dcpv}
\end{equation}
Results of the fit are given in Table~\ref{tab:kpp-dcpv}. 
The statistical significance of the asymmetry quoted in
Table~\ref{tab:kpp-dcpv} is calculated as
$\sqrt{-2\ln({\cal L}_0/{\cal L}_{\rm max})}$, where
${\cal L}_{\rm max}$ and ${\cal L}_0$ denote the maximum likelihood with the
best fit and with the asymmetry fixed at zero, respectively.
Systematic uncertainty for $\ACP$ results in Tbale~\ref{tab:kpp-dcpv} is 3\%.
%%%%%%%%%%%%%%%%%%%%%%%%%%%%%%%%%%%%%%%%%%%%%%%%%%%%%%%%%%%%%%%%%%%%%%%%%%%%%%%
\begin{figure}[!b]
 \begin{tabular}{lcr}
  \includegraphics[width=0.33\textwidth]{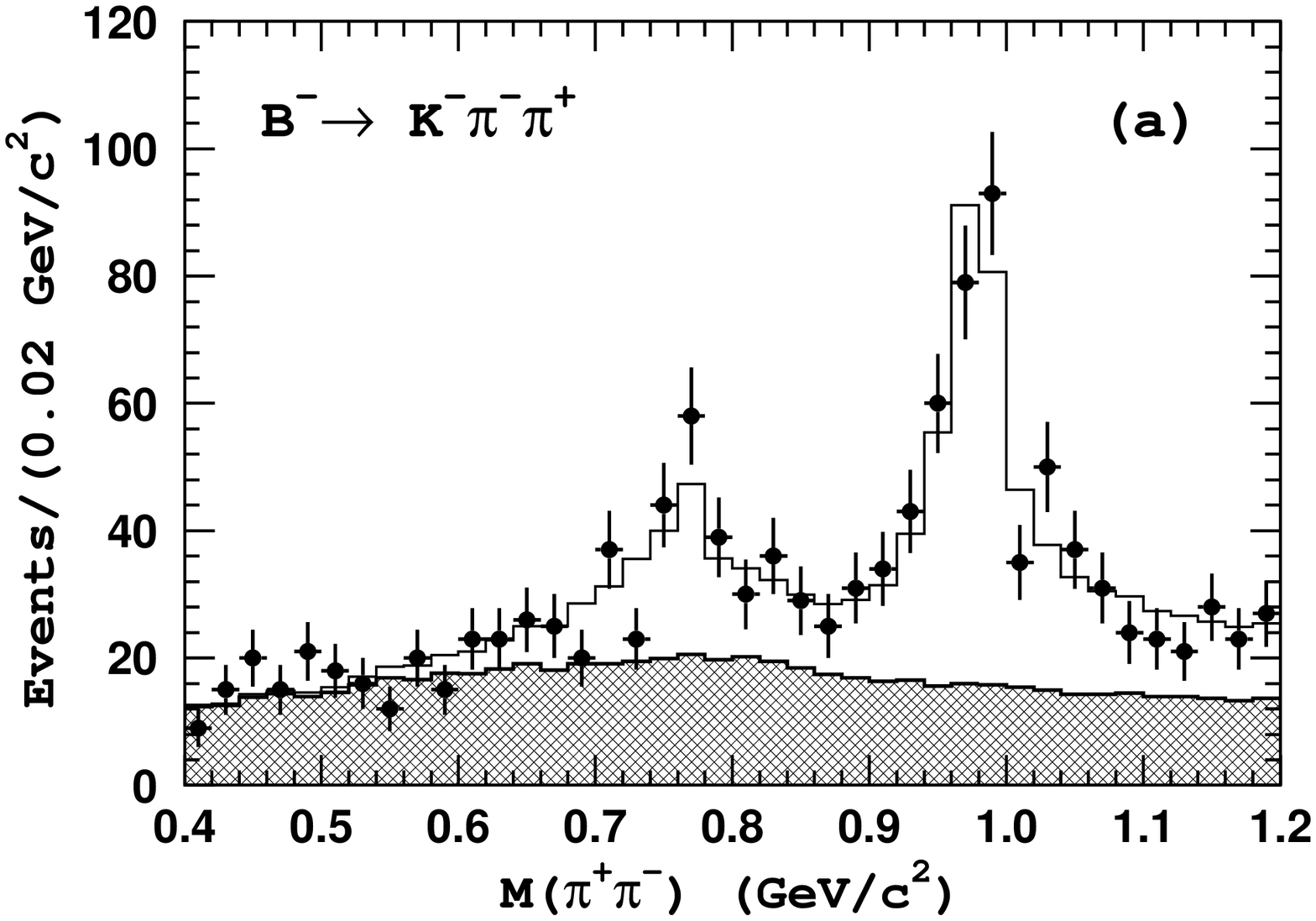}
  \includegraphics[width=0.33\textwidth]{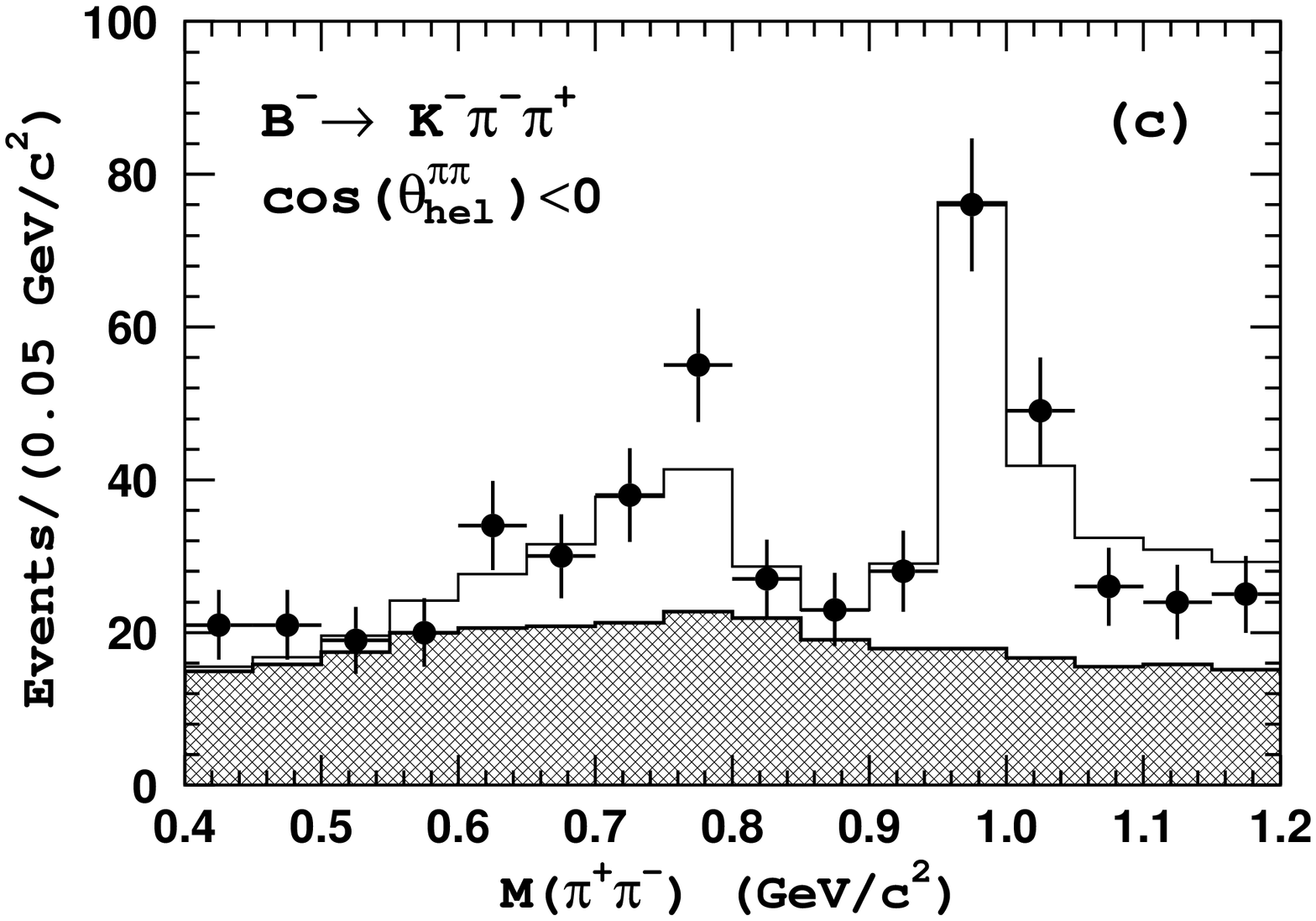}
  \includegraphics[width=0.33\textwidth]{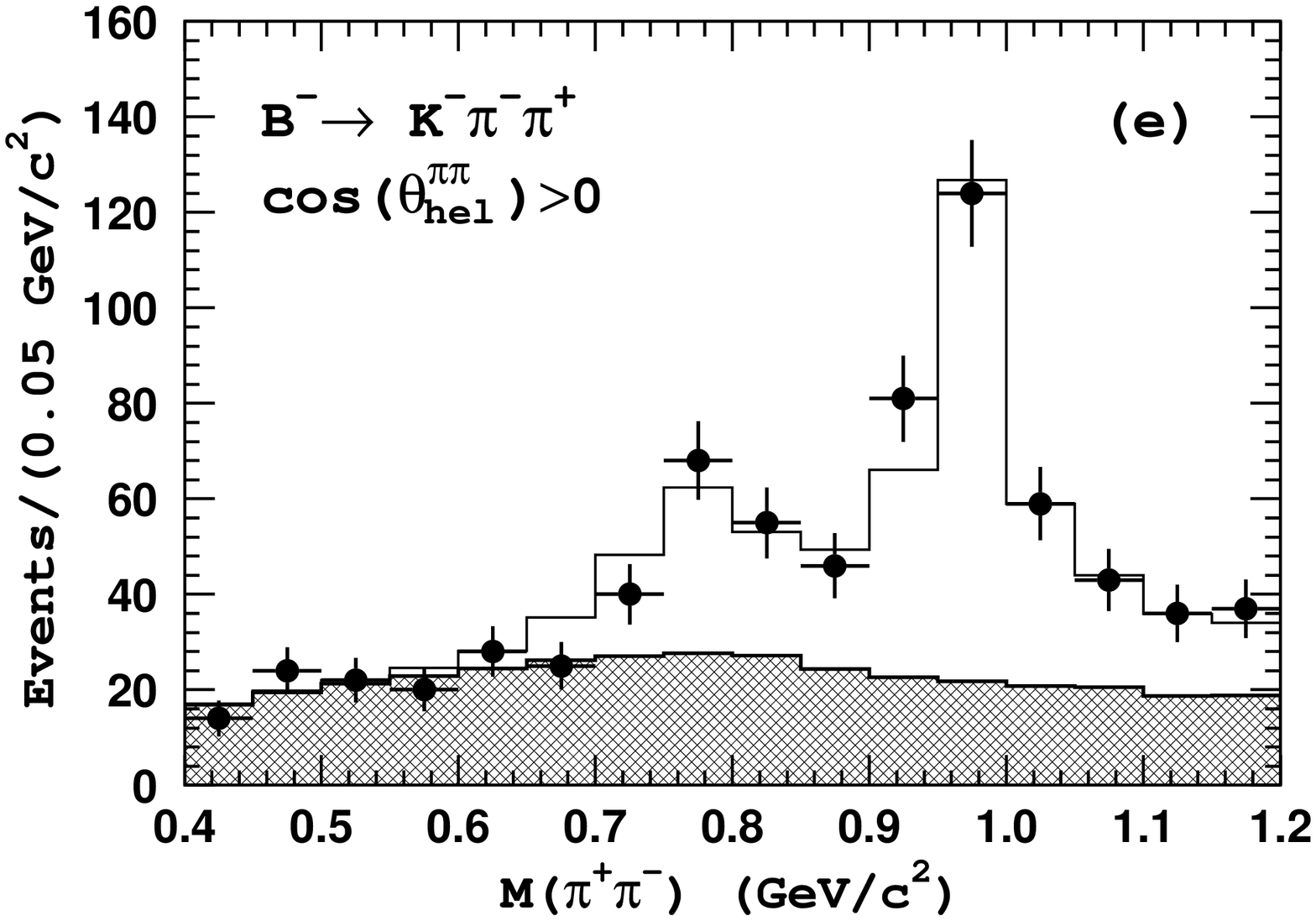}
\vspace*{-3mm} \\
  \includegraphics[width=0.33\textwidth]{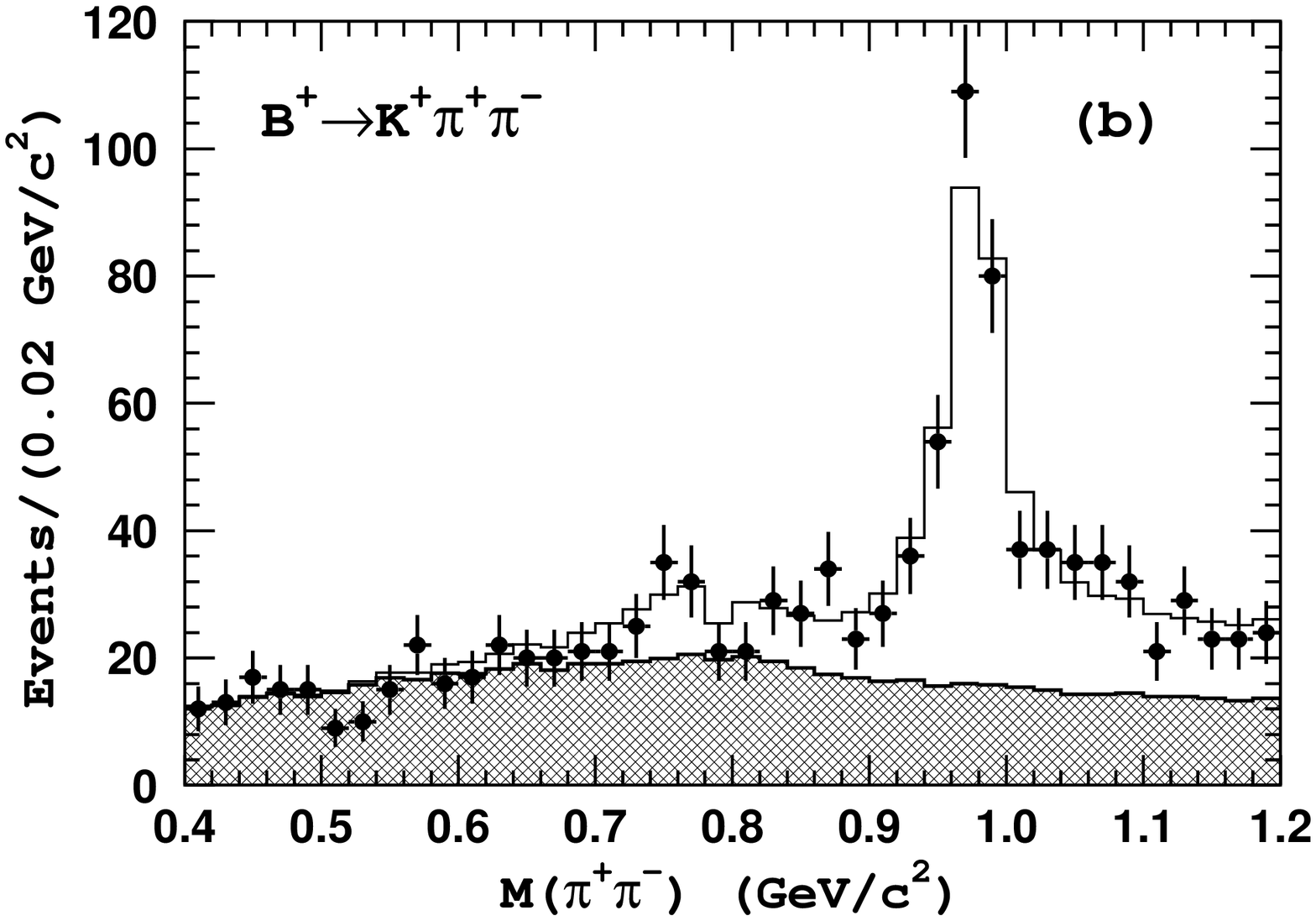}
  \includegraphics[width=0.33\textwidth]{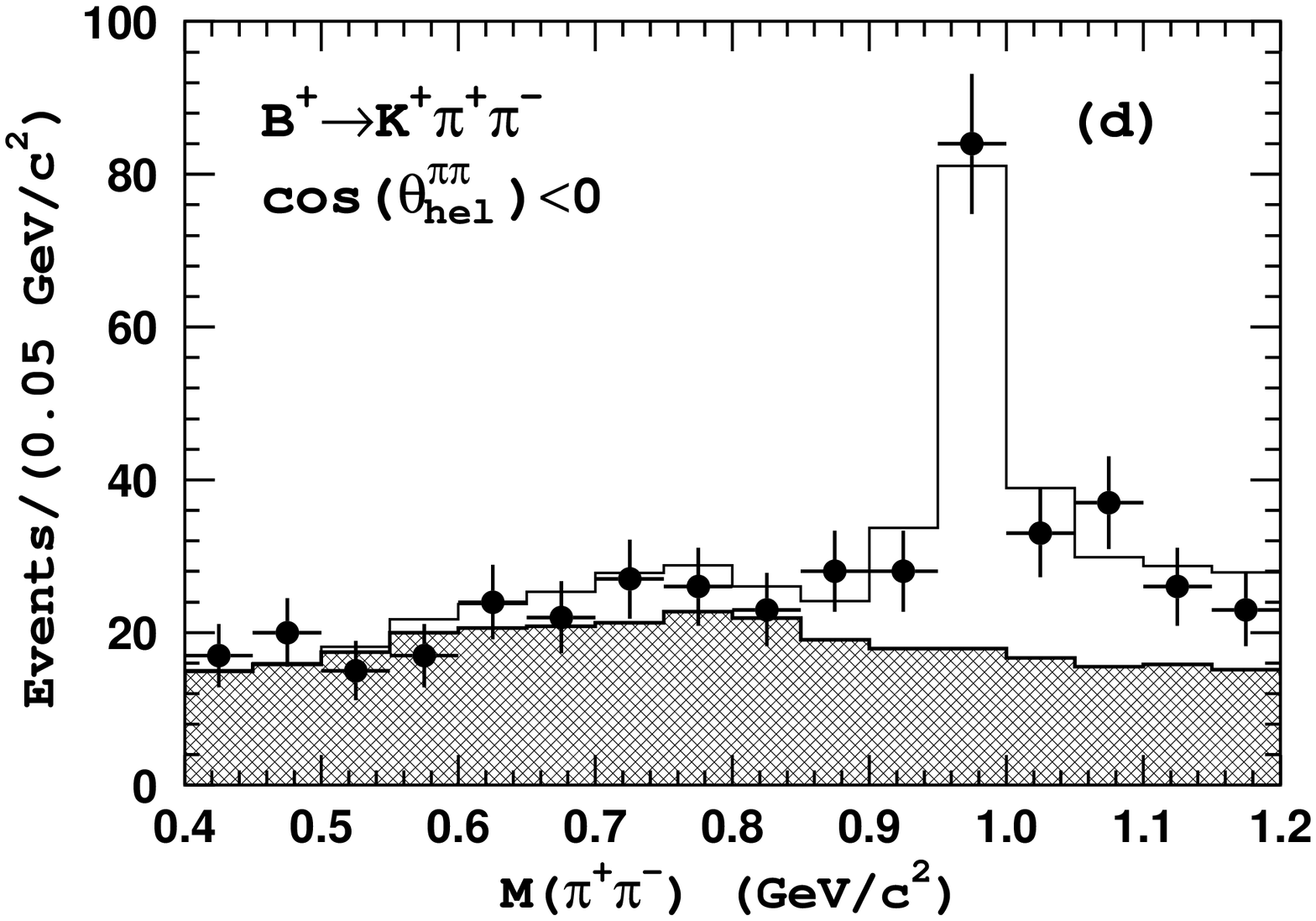}
  \includegraphics[width=0.33\textwidth]{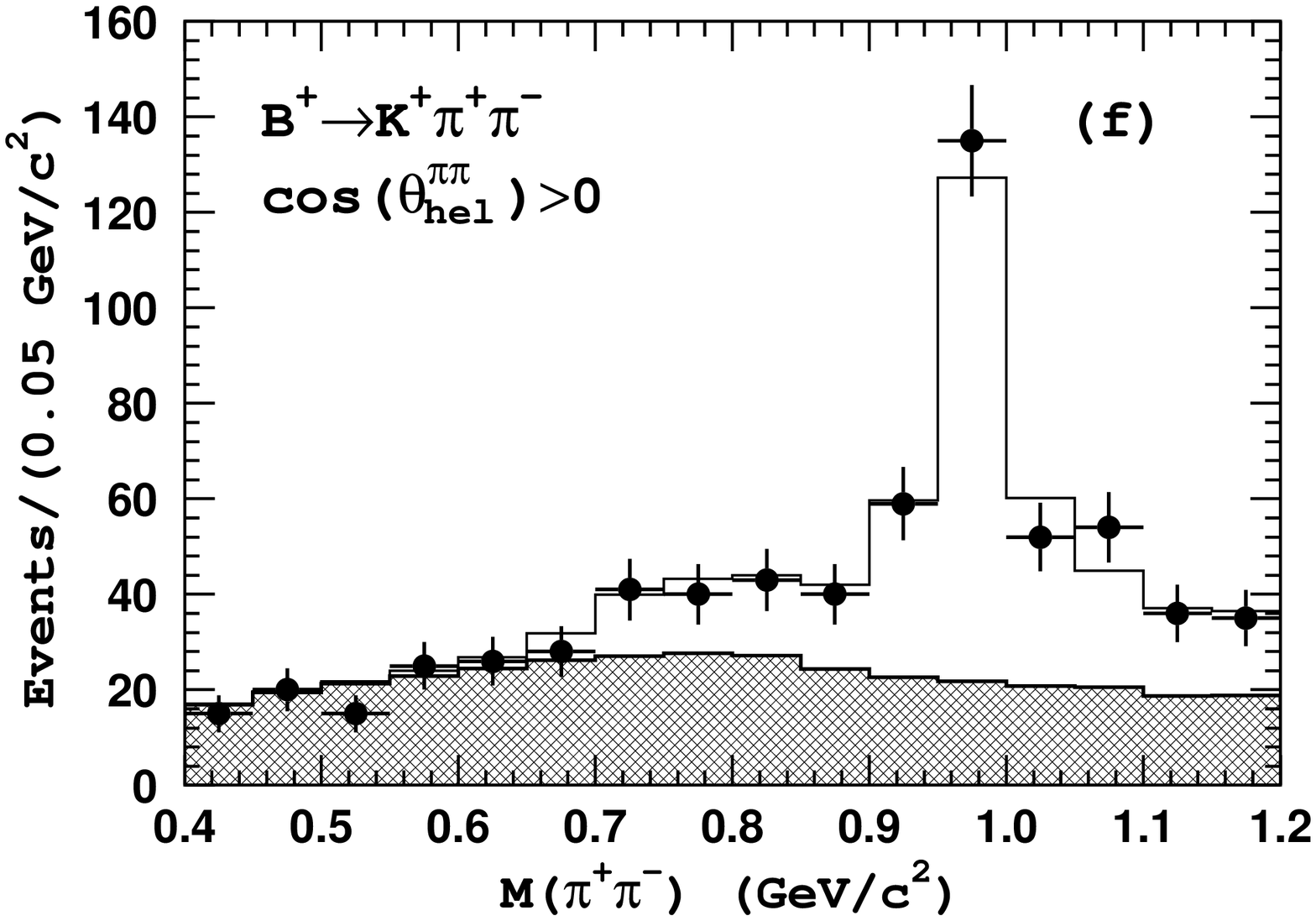}
\end{tabular}
\caption{$\pipi$ mass spectra for $B^-$ (top row) and $B^+$ (bottom row)
         events for different helicity regions:
         (a,b) no helicity cuts;
         (c,d) $\cos\theta_H^{\pi\pi}<0$;
         (e,f) $\cos\theta_H^{\pi\pi}>0$;
         Points with error bars are data, the open histogram is the fit
         result and the hatched histogram is the background component.}
\label{fig:kpp-dcpv}
\medskip
\end{figure}
%%%%%%%%%%%%%%%%%%%%%%%%%%%%%%%%%%%%%%%%%%%%%%%%%%%%%%%%%%%%%%%%%%%%%%%%%%%%%%%
The only
channel where the statistical significance of the asymmetry exceeds the
$3\sigma$ level is $B^\pm\to\rho(770)^0K^\pm$, where we find a $3.9\sigma$
effect. Figures~\ref{fig:kpp-dcpv}(a,b) show the $M(\pipi)$ distributions for
the $\rho(770)^0-f_0(980)$ mass region separately for $B^-$ and $B^+$ events.
The effect is more apparent when $M(\pipi)$ spectra for the two helicity
angle regions shown in Figs.~\ref{fig:kpp-dcpv}(c-f) are compared.
Results on the $\ACP$ measurement are summarized in Table~\ref{tab:kpp-dcpv}.
To cross check the asymmetry observed in $B^\pm\to\rho(770)^0K^\pm$ we make an
independent fit to $B^-$ and $B^+$ subsamples. We also confirm the significance
of the asymmetry observed in $B^\pm\to\rho(770)^0K^\pm$ channels with MC
pseudo-experiments. For more details see Ref.~\cite{kpp-dcpv-belle}. The
large value of $\ACP$ measured in $B^\pm\to\rho(770)^0K^\pm$ is in agreement
with a recent update by BaBar~\cite{kpp-dcpv-babar} and some theoretical
predictions~\cite{beneke-neubert}. The statistical significance of the
asymmetry varies from $3.7\sigma$ to $4\sigma$ depending
on the model used to fit the data. This is the first evidence for $CP$
violation in the decay of a charged meson.

\end{document}